\documentclass[12pt]{article}
\begin{document}

\begin{titlepage}

\begin{center}

\vspace{5mm}

\Large{\bf  Two-pion-exchange contributions to \\
the $pp\to pp\pi^0$ reaction }

\vspace{1cm}
  {\large Y. Kim$^{(a,b)}$,
  T. Sato$^{(c)}$, F.  Myhrer$^{(a)}$ and
  K. Kubodera$^{(a)}$}

\vskip 0.5cm
(a)~{\large Department of Physics and Astronomy,
University of South Carolina, Columbia,
South Carolina 29208, USA}

(b) {\large School of Physics,
Korea Institute for Advanced Study,
Seoul 130-012, Korea}

(c)   {\large  Department of Physics, Osaka University,
Toyonaka, Osaka 560-0043, Japan}

\end{center}


\vskip 1cm

\begin{abstract}
Our previous study of the near-threshold
$pp\to pp\pi^0$ reaction based
on a hybrid nuclear effective field theory
is further elaborated by examining the
momentum dependence of
the relevant transition operators.
We show that the two-pion exchange diagrams
give much larger contributions
than the one-pion exchange diagram,
even though the former is of higher order
in the Weinberg counting scheme.
The relation between our results and an alternative
counting scheme, the momentum counting scheme,
is also discussed.

\end{abstract}

\end{titlepage}

\newpage

In the standard nuclear physics approach (SNPA),
a nuclear reaction amplitude is calculated
with the use of the transition operator derived
from a phenomenological Lagrangian
and nuclear wave functions generated by a
high-precision phenomenological $NN$ potential.
SNPA has been enormously successful in
explaining a vast range of nuclear phenomena.
Meanwhile, a nuclear chiral perturbation approach based on
heavy-baryon chiral perturbation theory (HB$\chi$PT)
is gaining ground as a powerful tool
for addressing issues that cannot be readily settled in SNPA.
HB$\chi$PT is a low-energy effective field theory of QCD,
based on a systematic expansion in terms of
the expansion parameter $\epsilon\equiv Q/\Lambda_\chi \ll 1$,
where $Q$ is a typical energy-momentum involved
in a process under study or the pion mass $m_\pi$,
and the chiral scale
$\Lambda_\chi \simeq 4 \pi f_\pi \simeq 1$ GeV.
HB$\chi$PT has been applied with great success
to low-energy processes including
{\it e.g.,} pion-nucleon scattering and
electroweak reactions on a nucleon and in few-nucleon systems.
Our present work is concerned with a HB$\chi$PT study
of the near-threshold $pp$$\to$$pp\pi^0$ reaction.
A motivation of this study may be stated
in reference to the generic
$NN$$ \to$$NN \pi$  processes near threshold.
Although HB$\chi$PT presupposes the small size of its expansion
parameter  $Q/\Lambda_\chi$,
the pion-production reactions involve
somewhat large energy- and three-momentum transfers
even at threshold.
Therefore the application of HB$\chi$PT to the $NN$$\to$$NN \pi$  reactions
may involve some delicate aspects, but this also means that
these processes may serve as a good test case
for probing the limit of applicability of HB$\chi$PT.
Apart from this general issue to be investigated, a specific aspect of
the $pp$$\to$$pp\pi^0$ reaction makes its study particularly interesting.
For most isospin channels, the $NN$$\to$$NN\pi$ amplitude near threshold
is dominated by the pion rescattering diagram where the
$\pi N$ scattering vertex is given by
the Weinberg-Tomozawa term,
which represents the lowest chiral order contribution.
However, a quantitatively reliable description of  the $NN$$\to$$NN\pi$ reactions
obviously requires detailed examinations
of the corrections to this dominant amplitude.
Meanwhile, since the Weinberg-Tomozawa vertex does not
contribute to the pion-nucleon rescattering diagram
for $pp$$\to$$pp\pi^0$,
this reaction is particularly sensitive to higher chiral-order contributions
and hence its study is expected to provide valuable information to guide us
in formulating a quantitative description of  all the
$NN$$\to$$NN\pi$ reactions
(including the channels that involve a deuteron).

\vspace{3mm}

The first HB$\chi$PT-based study
of the near-threshold $pp$$\to$$pp\pi^0$ reaction
was made in Refs.~\cite{pmmmk96,cfmv96}.
In HB$\chi$PT one naturally expects
a small cross section for this reaction
since, for $s$-wave pion production,
the pion-nucleon vertex in the
impulse approximation (IA) diagram
and the pion-rescattering vertex in the
one-pion-exchange rescattering
(1$\pi$-Resc) diagram arise from the next-to-leading-order (NLO)
chiral lagrangian.
A remarkable feature found in Refs.~\cite{pmmmk96,cfmv96}
is that a drastic cancellation
between the IA  and 1$\pi$-Resc amplitudes
leads to the suppression of the $pp$$\to$$pp\pi^0$ amplitude
far beyond the above-mentioned naturally expected level.
This destructive interference is in sharp contrast
with the constructive interference reported
in SNPA-based calculations~\cite{kr66,ms91}.
It is to be recalled that the $pp$$\to$$pp\pi^0$  cross section
obtained in Refs.~\cite{kr66,ms91}
was significantly smaller (by a factor of $\sim$5)
than the experimental value~\cite{meyeretal90}.
The drastic cancellation between the IA and 1$\pi$-Resc terms
found in the HB$\chi$PT calculations~\cite{pmmmk96,cfmv96}
leads to even more pronounced disagreement between theory
and experiment.
In this connection it is worth noting that,
according to Lee and Riska~\cite{lr93},
the heavy-meson ($\sigma$ and $\omega$) exchanges
can strongly enhance the $pp$$\to$$pp\pi^0$ amplitude.
It is also to be noted that  $\sigma$-meson-exchange
introduced in many $NN$ potentials
is more properly described by correlated
two-pion-exchange
(see {\it e.g.,} Refs.~\cite{sb75,paris73}),
and that  there have been substantial developments  
in deriving a two-pion exchange $NN$ potential
using HB$\chi$PT, see e.g.~\cite{chiralNNpot}.
These developments were conducive to a HB$\chi$PT study of
two-pion-exchange (TPE) contributions to
the $pp$$\to$$pp\pi^0$ reaction~\cite{dkms99,am01}.
In the plane-wave approximation
it was found~\cite{dkms99} that
TPE contributions are indeed very large
(as compared to the 1$\pi$-Resc amplitude),
a result that is in line with the finding in Ref.\cite{lr93}.
A subsequent DWBA calculation~\cite{am01}
indicates that this feature remains essentially unchanged
when the initial- and final-state interactions 
are taken into account.
More recent investigations~\cite{hk02,hanhart04,Lensky05,hk07}, 
however,
have raised a number of important issues that 
call for further investigations,
and the purpose of our present note
is to address these issues.

\vspace{3mm}
In Ref.~\cite{dkms99}, to be referred to as DKMS, were derived
all the transition operators for $pp$$\to$$pp\pi^0$
belonging to next-to-next-to-leading order (NNLO)
in the Weinberg counting,
and these operators were categorized
into Types I $\sim$ VII, according to the patterns
of the corresponding Feynman diagrams;
see Figs. 2 - 5 in DKMS.
Types I, II, III and IV belong to diagrams of 
the two-pion exchange (TPE) type,
while Types V, VI and VII arise from diagrams 
of the vertex correction type.
A notable feature pointed out in DKMS is that
the contributions of Types II $\sim$ IV are by far the largest,
and that they even exceed those of the 1$\pi$-Resc amplitude,
which is formally of lower chiral order.
On the other hand,
the possibility of strong cancellation among the TPE diagrams
was pointed out in Refs.~\cite{hk02,hanhart04}.
This motivates us to make here a further study of
the behavior of the TPE diagrams.\footnote{For a brief report
on this study, see Ref.~\cite{fm06}.}

\vspace{3mm}
A remark is in order here on a counting scheme to be used.
At the $NN$$\to$$NN \pi$ threshold
the nucleon three-momentum must change from
the initial value $p \sim \sqrt{m_\pi m_N}$ to zero,
entailing a rather large momentum transfer.
To take this large momentum transfer into account,
Cohen {\it et al.}~\cite{cfmv96} proposed a new counting scheme,
to be called the momentum counting scheme (MCS);
see Ref.~\cite{hanhart04} for a detailed review.
In MCS the expansion parameter is
$\tilde{\epsilon}\equiv p/m_N \simeq (m_\pi / m_N)^{1/2}$,
which is larger than the usual HB$\chi$PT expansion parameter
$\epsilon \simeq m_\pi / m_N$.
A study based on MCS~\cite{hanhart04} indicates
that the 1$\pi$-Resc diagram for $pp$$\to$$pp\pi^0$
is higher order in $\tilde{\epsilon}$ (and hence less important)
than a certain class of TPE diagrams, called
``leading order loop diagrams",
and that MCS is consistent with the estimates
of the TPE and other diagrams reported in DKMS.
Furthermore, according to Hanhart and Kaiser (HK)~\cite{hk02},
the ``leading parts" (see below) of these MCS
``leading order" diagrams
exhibit exact cancellation among themselves;\footnote{HK~\cite{hk02}
pointed out that the sign of the contribution of Type II
in Ref.~\cite{dkms99} should be reversed;
we have confirmed the necessity of this correction.}
see also Lensky {\it et al.}~\cite{Lensky05}.
Although these studies are illuminating,
we consider it important to examine the behavior of
the ``sub-leading" parts (in MCS counting)
of these TPE diagrams in order to see whether
they can be still as large as indicated
by the phenomenological success of the Lee-Riska
heavy-meson exchange mechanism.
In what follows we shall demonstrate that this
is indeed the case.

\vspace{3mm}
Analytic expressions for the $pp$$\to$$pp\pi^0$
transition operators to NNLO in HB$\chi$PT
were given in DKMS.
Although these expressions are valid for arbitrary kinematics,
we find it illuminating to concentrate here
on their simplified forms obtained with the use of
fixed kinematics approximation (FKA),
wherein the energies associated with particle propagators are
``frozen" at their threshold values.
In FKA, the TPE operator corresponding to each
of the above-mentioned Types I $\sim$ IV can be written as:
\begin{eqnarray}
T &=& \left( \frac{g_A}{f_\pi} \right)
\left( \vec{\Sigma}\cdot \vec{k} \right)
t(p,p^\prime , x)
\label{eq:T}
\end{eqnarray}
where $\vec{p}$ ($\vec{p}^{\; \prime}$)
is the relative three-momentum
in the initial (final) $pp$ state
($\vec{p}_1 - \vec{p}_2 = 2\vec{p}$,
$\vec{p}_1^{\; \prime} - \vec{p}_2^{\; \prime}
= 2\vec{p}^{\; \prime}$),
$\vec{k}\equiv\vec{p}-\vec{p}^{\; \prime}$,
$x=\hat{p}\cdot\hat{p}^\prime$, and
$\vec{\Sigma} = \frac{1}{2}
(\vec{\sigma}_1-\vec{\sigma}_2)$.
The function $t(p,p^\prime,x)$ diverges as $k\to\infty$,
and it is useful to decompose $t(p,p^\prime,x)$
into terms that have definite $k$-dependence as $k\to\infty$.
It turns out~\cite{sm99} that $t(p,p^\prime,x)$ can be expressed as
\begin{eqnarray}
t(p,p^\prime , x) \stackrel{k\to\infty}{\sim} \;
\;t_1\! \left( g_A/(8f_\pi^2)\right)^2| \vec{k} |
+ t_2\! \left(\ln\{|\vec{k} |^2/\Lambda^2\} \right)
+\, t_3 + \delta t(p,p^\prime , x),
\label{eq:tasympt}
\end{eqnarray}
where $t_3$ is asymptotically $k$-independent,
and $\delta t(p,p^\prime , x) $ is  ${\cal O}(k^{-1})$.
For each of Types I $\sim$ IV,
analytic expressions for $t_i$'s ($i=1, 2, 3$)
can be extracted~\cite{sm99} from
the amplitudes $T$ given in DKMS~\cite{dkms99}.
The first term with $t_1$ in eq.(\ref{eq:tasympt}) is
the leading part in MCS discussed by HK~\cite{hk02},
whereas the remaining terms,
which we refer to as the ``sub-leading" terms,
were not considered by HK.
The study of these sub-leading terms is an important theme
in what follows.
Table 1 shows the value of $t_1$  for Type K
(K= I $\sim$ IV) extracted from the results given in DKMS.
The third row in Table 1 gives the ratio
$R_K = T_K /T_{Resc}$, where $T_K$ is
the plane-wave matrix element of $T$
in eq.(\ref{eq:T}) for Type K (K=I $\sim$ IV) normalized
by $T_{Resc}$, the plane-wave matrix element of
the 1$\pi$-Resc diagram.
The fourth row in Table 1
gives $R^{\,\star}_K = T^{\,\star}_K /T_{Resc}$,
where $T_K^{\,\,\star}$ is
the plane-wave matrix element of $T$
with the $t_1$ term in eq.(\ref{eq:tasympt}) subtracted.
We can see from the table that the most divergent $t_1$ terms
of the TPE diagrams add up to zero,
confirming the result of Ref.~\cite{hk02}.
However, this does not necessarily mean
that the TPE diagrams are unimportant, because
we still need to examine the contributions
of the ``sub-leading" terms
(the $t_2$, $t_3$ and $\delta t$ terms) in eq.(\ref{eq:tasympt}).
Comparison of $R_K$ and $R^{\,\star}_K$ indicates
that the subtraction of the $t_1$ term
reduces the magnitude of $T_K$ drastically
(except for Type I which has no $t_1$ term),
but the fact that $|R^{\,\star}_K|$ is of the order of
unity (Types I, II and IV) or larger than 1 (Type III)
suggests that the TPE contributions can be quite important.
The sum of the contributions of Types I $\sim$ IV is
\begin{eqnarray}
\sum_K R^{\,\star}_K\;\; (=\sum_K R_K)\;=\;-4.65\,,
\end{eqnarray}
which indicates that, at least in plane-wave approximation,
the TPE contributions are more important than
the 1$\pi$-Resc contribution.

\begin{center}
 \parbox[t]{5.3in}{ Table 1:
For the four types of TPE diagrams,
K= I, II, III and IV,
the second row gives the value of $t_1$ defined in eq.(\ref{eq:tasympt}),
and the third row gives the ratio
$R_K = T_K /T_{Resc}$, where $T_K$ is
the plane-wave matrix element of
$T$ in eq.(\ref{eq:T}) for Type K,
and $T_{Resc}$ is the 1$\pi$-Resc amplitude.
The last row gives
$R_K^{\,\,\star} = T_K^{\,\,\star} /T_{Resc}$,
where $T_K^{\,\,\star}$ is
the plane-wave matrix element of
$T$ in eq.(\ref{eq:T}) with the $t_1$ term in eq.(\ref{eq:tasympt})
subtracted.}
\end{center}
$$
\begin{array}{|l| |r| r|r|r|r|}
\hline
{\rm Type\; of\; diagrams:  K=} & {\rm I}&{\rm II}&{\rm III}&{\rm IV} \\
\hline
 (t_1)_K \; \; \; \; \; & 0 & 1  & 1/2 & - 3/2  \\
\hline
 R_K  & -.70& -6.54& -6.60 & 9.19  \\
\hline
R^{\,\star}_K& -.70& -0.82& -3.73& 0.61\\
\hline
\end{array}
$$

\vspace{6mm}
Next we investigate the behavior of the TPE diagrams
as we go beyond the plane-wave approximation by using
distorted waves (DW) for the initial- and final-state
$NN$ wave functions.
For formal consistency
we should use the $NN$ potential derived from HB$\chi$PT,
but we adopt here a ``hybrid EFT" approach
and use phenomenological potentials.
A conceptual problem in adopting this hybrid approach
is that, whereas the TPE transition operators
derived in HB$\chi$PT are valid only for a
momentum range sufficiently lower than $\Lambda_\chi$$\sim$1 GeV,
a phenomenological $NN$ potential
can in principle contain any momentum components.\footnote{
A pragmatic problem associated with this conceptual issue
is that, in a momentum-space calculation
of the matrix elements of the TPE operators sandwiched between
distorted $pp$ wave-functions generated by a phenomenological
$NN$ potential,
the convergence of momentum integrations
is found to be extremely slow~\cite{slmk,sm99}.}
To stay close to the spirit of  HB$\chi$PT,
we therefore introduce a Gaussian momentum regulator,
$\exp(-p^2/\Lambda_G^2)$, in the initial and final
distorted wave integrals, suppressing thereby
the high momentum components of
the phenomenological $NN$ potentials;
this is similar to the MEEFT method used in Ref.\cite{TSPark}.
$\Lambda_G$ should be larger than the characteristic
momentum scale of the $pp$$\to$$pp\pi^0$ reaction,
$p \simeq \sqrt{m_N m_\pi} \simeq 360$ MeV/c,
but it should not exceed the chiral scale $\Lambda_\chi$;
in the present study we shall consider the range,
500 MeV$<\Lambda_G<$1 GeV.
As high-precision phenomenological $NN$ potentials,
we consider the Bonn-B potential~\cite{BonnB}, the 
CD-Bonn potential~\cite{CDBonn}, and the Nijm93 potential 
of the Nijmegen group~\cite{Nijmegen}.

\vspace{3mm} 

It is worth noting here that several groups 
~\cite{EGM,Vlowk} have developed
a systematic approach to construct from a phenomenological potential
an effective $NN$ potential,  
called V$_{low-k}$,
that resides within a model space
which only contains momentum components below a specified
cutoff scale $\Lambda_{low-k}$. 
In this work we will use V$_{low-k}$ as 
derived by the Stony Brook group~\cite{Vlowk}. 
It is conceptually natural to use V$_{low-k}$
in conjunction with transition operators derived
from HBChPT~\cite{Kimetal06}.
A problem however is 
V$_{low-k}$~\cite{Vlowk},
primarily meant for describing sub-pion-threshold phenomena,
was obtained with the use of a rather low cutoff,
$\Lambda_{low-k}\sim$ 2 fm$^{-1}$.
This cutoff is perhaps too close to the characteristic momentum scale
$p\sim$ 360 MeV/$c$ for the pion production reaction.
It therefore seems worthwhile to
 ``rederive" V$_{low-k}$
employing a momentum cut-off
higher than 2 fm$^{-1}$
and use it in the present DWBA calculation.
Below we will use V$_{low-k}$  generated
from the CD-Bonn potential
for $\Lambda_{low-k}$= 4 and 5 fm$^{-1}$.
We remark that, as is well known,
V$_{low-k}$'s generated from any realistic phenomenological
potentials lead to practically equivalent half-off-shell $NN$ K-matrices
and hence the same NN wave function.

\vspace{3mm} 

We evaluate the TPE contributions in DWBA
for a typical case of $T_{lab} = 281 $ MeV.
Since the $t_1$ terms in eq.(\ref{eq:tasympt}) add up to zero,
we drop the $t_1$ terms in our calculation.\footnote{
Removing the $t_1$ term lessens
the severity of the convergence problem
in our momentum integration mentioned
in footnote 3.}
Thus, in eq.(1), we use $t^\star(p,p',x)$
instead of $t(p,p',x)$,
where $t^\star(p,p',x)$ is obtained from $t(p,p',x)$
by suppressing the $t_1$ term.
The partial-wave projected form of
$t^\star(p,p^\prime,x)$ in a DWBA calculation
is written as:
\begin{eqnarray}
J&=& -\left(\frac{m_Nm_\pi}{8\pi}\right)\;
\int_0^\infty p^2 {\rm d}p\; p^{\prime\; 2}
{\rm d}p^\prime
\int_{-1}^1 {\rm d}x\; \psi_{^1\!S_0}(p^\prime )\;
t^\star(p,p^\prime,x) \; (p-p^\prime x) \psi_{^3\!P_0}(p)
\label{eq:J}
\end{eqnarray}
Here $\psi_\alpha (p)$ is
a distorted two-nucleon relative wave function
in the $\alpha$ partial-wave
($^1\!S_0$ for the initial state and $^3\!P_0$
for the final state) given by
\begin{eqnarray}
\psi_\alpha (p) &=& \cos(\delta_\alpha ) \; \Big[
\delta (p- p_{on} ) / p^2 +
{\cal P}\left(\frac{K_\alpha (p, p_{on} ) }{(E-E_p ) }\right)
\Big] \; ,
\end{eqnarray}
where
$\delta_\alpha$ is the phase-shift for the $\alpha$ partial wave,
and $K_\alpha(p,p_{on})$ is the partial-wave K-matrix
pertaining to the asymptotic on-shell momentum $p_{on}$.
The plane-wave approximation corresponds to
the use of the wave functions of the generic form:
\begin{eqnarray}
\psi(p) &=&
\delta (p- p_{on} ) / p^2  \; .
\end{eqnarray}

\vspace{4mm}
We show in Table 2 the values of $J$, eq.(\ref{eq:J}),
for the TPE operators of Types I $\sim$ IV,
calculated at $T_{lab}= 281$ MeV,
with the use of  the Nijm93 potential 
of the Nijmegen group~\cite{Nijmegen}\footnote{
We  have checked the results obtained using 
the Bonn-B and CD-Bonn $NN$
potentials are very similar to those for the 
Nijm93 
potential case,
which we show here as a representative case. }
and $V_{low-k}$.
For the Nijm93 potential case,
we present the results for five different values of $\Lambda_G$
between 500 and 1000 MeV/$c$.
For the $V_{low-k}$ case,
the results for two choices of $\Lambda_{low-k}$
are shown: $\Lambda_{low-k}$ = 4 fm$^{-1}$ and 5 fm$^{-1}$.
For comparison, the values of $J$  corresponding to plane-wave
approximation are also shown (bottom row).
From Table 2 we learn the following:
(1) The results for the Nijm93 potential 
with the gaussian cutoff
$\Lambda_G$ are stable against the variation of $\Lambda_G$
within a reasonable range (500 - 1000 MeV/c);
(2) There is semi-quantitative agreement between the results
for the Nijm93 potential and those for $V_{low-k}$;
(3) A semi-quantitative agreement is also seen between
the DWBA and PWBA calculations;
(4) The feature found in the plane-wave approximation
that the contributions of the TPE diagrams
are more important than the 1$\pi$-Resc contribution
remains unchanged in the DWBA calculation;
the summed contribution of the TPE operators
is larger (in magnitude) than that of 1$\pi$-Resc
by a factor of 2$\sim$3.5.

\vspace{4mm}
\begin{center}
 \parbox[t]{5.3in}{ Table 2:
The values of $J$, eq.(\ref{eq:J}), corresponding to
the TPE diagrams of Types I $\sim$ IV,
evaluated in a DWBA calculation for $T_{lab}= 281$ MeV.
The column labeled ``Sum" gives
the combined contributions of Types I $\sim$ IV,
and the last column gives the value of $J$ for 1$\pi$-Resc.
For the Nijm93 potential case,
the results for five different choices of  $\Lambda_G$
are shown.
For the case with V$_{low-k}$,
CD-4 (CD-5) represents V$_{low-k}$
generated from the CD-Bonn potential
with a momentum cut-off
$\Lambda_{low-k}$ = 4\,fm$^{-1}$ (5\,fm$^{-1}$).
The last row gives the results obtained 
in plane-wave approximation.
}
\end{center}
$$
\begin{array}{|l||r|r|r|r|c||r| }
\hline
    &\!\! {\rm I}&{\rm II}\; \; &{\rm III}\; \; &{\rm IV} \; \; &
\;\:\:\;{\rm Sum} \:\:\; & 1\pi\!\!-\!\!{\rm Resc} \\
\hline
V_{\rm Nijm}\,:\Lambda_G=500 \,{\rm MeV}/c&  -0.11 & -0.12 &  -0.55
& 0.08 & -0.70 &  0.18\\
\hline
V_{\rm Nijm}\,:\Lambda_G=600 \,{\rm MeV}/c& -0.12 & -0.12 &  -0.57
& 0.07 & -0.74 &0.20 \\
\hline
V_{\rm Nijm}\,:\Lambda_G=700 \,{\rm MeV}/c& -0.12 & -0.11 &  -0.57
& 0.06 & -0.74 & 0.21\\
\hline
V_{\rm Nijm}\,:\Lambda_G=800 \,{\rm MeV}/c& -0.12 & -0.11 &  -0.55
& 0.04 & -0.74 & 0.22 \\
\hline
V_{\rm Nijm}\,:\Lambda_G=1000 \,{\rm MeV}/c& -0.12  & -0.10 &  -0.52
& 0.03 & -0.71 &0.23 \\
\hline \hline
V_{low-k}  \; \; (CD\!-\!4) & -0.12  & -0.09&  -0.46
& 0.03 & -0.65 & 0.23  \\
\hline
V_{low-k}  \; \; (CD\!-\!5) & -0.09  & -0.06&  -0.30
& -0.01 & -0.46 & 0.22 \\
\hline \hline
{\rm Plane\!-\!wave} & -0.06   & -0.07 &  -0.30
& 0.05& -0.37 &  0.080 \\
\hline
\end{array}
$$

\vspace{10mm}
We now discuss the above results
in the context of MCS~\cite{hanhart04}.
A subtlety in MCS is that a loop diagram of a given order
$\nu$  in $\tilde \epsilon$
not only contains a contribution of order $\nu$
(``leading part")
but, in principle, can also involve contributions
of higher orders in $\tilde \epsilon$
(``sub-leading part")
due to the non-analytic functions generated
by the loop integral.
As mentioned, however, HK~\cite{hk02}
considered only the leading part,
which correspond to the $t_1$ term
in eq.(\ref{eq:tasympt}).
According to MCS, for the reaction $pp\to pp \pi^0$,
the loop diagrams corresponding to our
Type II, III and IV diagrams belong to NLO in the 
$\tilde \epsilon$ parameter, 
whereas those corresponding to Type I
and the 1$\pi$-Resc tree diagram
are next order in $\tilde \epsilon$ (NNLO);
see Table 11 in Ref.~\cite{hanhart04}.
Meanwhile, as discussed earlier,
the sum of the ``leading parts" of the 
NLO diagrams vanishes,
and therefore, in calculating $J$'s in Table 2,
we have dropped the $t_1$ term contribution,
retaining only the ``sub-leading" parts of 
these NLO diagrams.
This means that all the entries in Table 2
represent ``sub-leading contributions" (NNLO) in MCS.
If we look at Table 2 from this perspective,
we note that the order-of-magnitude behavior
of our numerical results is in rough agreement with MCS,
although Type IV tends to be
rather visibly smaller (in magnitude)
than the others.
However, it is striking that $J$ for Type III
is significantly (if not by an order of magnitude)
larger than the other sub-leading contributions.
(A similar feature was also seen in $R^{\star}$ in Table 1.)
In view of the fact that Type III
arises from crossed-box TPE diagrams~\cite{dkms99},
there is a possibility that the enhancement
of the Type III diagrams may be related to the
strong attractive scalar $NN$ potential that is known
to arise from TPE crossed-box-diagrams~\cite{sb75,paris73}.

\vspace{3mm}
We have studied the ``sub-leading" parts, 
which are of NNLO  in the momentum counting
scheme (MCS)~\cite{hanhart04},
of the TPE amplitudes 
for the $pp$$\to$$pp\pi^0$ reaction
in both PWBA and DWBA calculations.  
We have shown in 
fixed kinematics approximation (FKA) that,
even though the leading parts of the
TPE amplitudes cancel among themselves~\cite{hk02,Lensky05},
the contributions of the sub-leading parts
are quite significant.
They are in general comparable to the 1$\pi$-Resc amplitude,
and the sub-leading part of the Type III diagrams
is even significantly larger than the 1$\pi$-Resc diagram.
The total contribution of the TPE diagrams
is larger (in magnitude) than that of the 1$\pi$-Resc diagram
by a factor of $\sim$5 (PWBA) or 2$\sim$3 (DWBA).
We have focused here on the TPE loop diagrams
but,  to obtain theoretical cross section
for $pp$$\to$$pp\pi^0$ that can be directly compared
with the experimental value,
we must consider the other diagrams discussed in DKMS
as well as the relevant counter terms.
These will be discussed in a forthcoming article~\cite{kkms07} .

\vspace{3mm}

The authors are indebted to Christoph Hanhart and Anders G{\aa}rdestig
for useful discussions. 
A helpful communication from Ulf Meissner is 
also grateful acknowledged.
This work is supported in part
by the US National Science Foundation,
Grant No.  PHY-0457014, and
by the Japan Society for the Promotion of Science,
Grant-in-Aid for Scientific Research (C) No.15540275
and Grant-in-Aid for Scientific Research on Priority Areas (MEXT),
No. 18042003.

\end{document}